\documentclass[aps,prl,preprint,tightenlines,superscriptaddress,showpacs,byrevtex]{revtex4}

\def \ks {K_{\rm S}^0}
\usepackage{graphicx}
\usepackage{color}

\begin{document}

% --------------------------------------------------
\preprint{\vbox{
\hbox{Belle Preprint 2010-2}
\hbox{KEK   Preprint 2009-41}
\hbox{NTLP Preprint 2010-01}
}}

\title{ \quad\\[0.5cm] Search for Lepton Flavor Violating $\tau^-$ Decays\\
into {$\ell^-\ks$} and $\ell^-\ks\ks$}
\begin{abstract}
We have searched for 
{the {lepton-flavor-violating} {decays}}
$\tau^-\rightarrow \ell^-\ks$ 
and $\ell^-\ks\ks$ ($\ell = e \mbox{ or } \mu$),
using a data sample of
671 fb$^{-1}$ collected with
the Belle detector at the 
{KEKB asymmetric-energy $e^+e^-$ collider.}
No evidence for a signal {was} found
in {any} of the decay modes,
{and we} set the following upper limits 
for the branching {fractions:}
${\cal{B}}(\tau^-\rightarrow e^-\ks) < 2.6\times 10^{-8}$,
${\cal{B}}(\tau^-\rightarrow \mu^-\ks) < 2.3\times 10^{-8}$, 
${\cal{B}}(\tau^-\rightarrow e^-\ks\ks) < 7.1\times 10^{-8}$
and
${\cal{B}}(\tau^-\rightarrow \mu^-\ks\ks) < 8.0\times 10^{-8}$ 
at the 90\% confidence level. 
\end{abstract}
\affiliation{Budker Institute of Nuclear Physics, Novosibirsk, Russian Federation}
\affiliation{Faculty of Mathematics and Physics, Charles University, Prague, The Czech Republic}
\affiliation{University of Cincinnati, Cincinnati, OH, USA}
\affiliation{Department of Physics, Fu Jen Catholic University, Taipei, Taiwan}
\affiliation{The Graduate University for Advanced Studies, Hayama, Japan}
\affiliation{Hanyang University, Seoul, South Korea}
\affiliation{University of Hawaii, Honolulu, HI, USA}
\affiliation{High Energy Accelerator Research Organization (KEK), Tsukuba, Japan}
\affiliation{Institute of High Energy Physics, Chinese Academy of Sciences, Beijing, PR China}
\affiliation{Institute for High Energy Physics, Protvino, Russian Federation}
\affiliation{Institute of High Energy Physics, Vienna, Austria}
\affiliation{INFN - Sezione di Torino, Torino, Italy}
\affiliation{Institute for Theoretical and Experimental Physics, Moscow, Russian Federation}
\affiliation{J. Stefan Institute, Ljubljana, Slovenia}
\affiliation{Kanagawa University, Yokohama, Japan}
\affiliation{Institut f\"ur Experimentelle Kernphysik, Karlsruhe Institut f\"ur Technologie, Karlsruhe, Germany}
\affiliation{Korea Institute of Science and Technology Information, Daejeon, South Korea}
\affiliation{Korea University, Seoul, South Korea}
\affiliation{Kyungpook National University, Taegu, South Korea}
\affiliation{\'Ecole Polytechnique F\'ed\'erale de Lausanne, EPFL, Lausanne, Switzerland}
\affiliation{Faculty of Mathematics and Physics, University of Ljubljana, Ljubljana, Slovenia}
\affiliation{University of Maribor, Maribor, Slovenia}
\affiliation{Max-Planck-Institut f\"ur Physik, M\"unchen, Germany}
\affiliation{University of Melbourne, Victoria, Australia}
\affiliation{Nagoya University, Nagoya, Japan}
\affiliation{Nara Women's University, Nara, Japan}
\affiliation{National Central University, Chung-li, Taiwan}
\affiliation{National United University, Miao Li, Taiwan}
\affiliation{Department of Physics, National Taiwan University, Taipei, Taiwan}
\affiliation{H. Niewodniczanski Institute of Nuclear Physics, Krakow, Poland}
\affiliation{Nippon Dental University, Niigata, Japan}
\affiliation{Niigata University, Niigata, Japan}
\affiliation{University of Nova Gorica, Nova Gorica, Slovenia}
\affiliation{Novosibirsk State University, Novosibirsk, Russian Federation}
\affiliation{Osaka City University, Osaka, Japan}
\affiliation{Panjab University, Chandigarh, India}
\affiliation{Peking University, Beijing, PR China}
\affiliation{Saga University, Saga, Japan}
\affiliation{University of Science and Technology of China, Hefei, PR China}
\affiliation{Seoul National University, Seoul, South Korea}
\affiliation{Sungkyunkwan University, Suwon, South Korea}
\affiliation{School of Physics, University of Sydney, NSW 2006, Australia}
\affiliation{Tata Institute of Fundamental Research, Mumbai, India}
\affiliation{Toho University, Funabashi, Japan}
\affiliation{Tohoku Gakuin University, Tagajo, Japan}
\affiliation{Tohoku University, Sendai, Japan}
\affiliation{Department of Physics, University of Tokyo, Tokyo, Japan}
\affiliation{Tokyo Metropolitan University, Tokyo, Japan}
\affiliation{Tokyo University of Agriculture and Technology, Tokyo, Japan}
\affiliation{IPNAS, Virginia Polytechnic Institute and State University, Blacksburg, VA, USA}
\affiliation{Yonsei University, Seoul, South Korea}
\author{Y.~Miyazaki} % Nagoya 
\affiliation{Nagoya University, Nagoya, Japan}
\author{I.~Adachi} % KEK 
\affiliation{High Energy Accelerator Research Organization (KEK), Tsukuba, Japan}
\author{H.~Aihara} % Tokyo 
\affiliation{Department of Physics, University of Tokyo, Tokyo, Japan}
\author{K.~Arinstein} % BINP  Novosibirsk 
\affiliation{Budker Institute of Nuclear Physics, Novosibirsk, Russian Federation}
\affiliation{Novosibirsk State University, Novosibirsk, Russian Federation}
\author{V.~Aulchenko} % BINP  Novosibirsk 
\affiliation{Budker Institute of Nuclear Physics, Novosibirsk, Russian Federation}
\affiliation{Novosibirsk State University, Novosibirsk, Russian Federation}
\author{A.~M.~Bakich} % Sydney 
\affiliation{School of Physics, University of Sydney, NSW 2006, Australia}
\author{V.~Balagura} % ITEP 
\affiliation{Institute for Theoretical and Experimental Physics, Moscow, Russian Federation}
\author{E.~Barberio} % Melbourne 
\affiliation{University of Melbourne, Victoria, Australia}
\author{A.~Bay} % Lausanne 
\affiliation{\'Ecole Polytechnique F\'ed\'erale de Lausanne, EPFL, Lausanne, Switzerland}
\author{K.~Belous} % Protvino 
\affiliation{Institute for High Energy Physics, Protvino, Russian Federation}
% ...
\collaboration{The Belle Collaboration}
\noaffiliation

\pacs{11.30.Fs; 13.35.Dx; 14.60.Fg}

\maketitle

\section{Introduction}

{Lepton flavor violation (LFV)
in charged lepton decays is forbidden 
{in the Standard Model (SM)
or highly suppressed} if neutrino mixing is included.
However, LFV {appears} in {various} extensions of the SM.
{In particular, 
{the}
{lepton-flavor-violating decays}
 $\tau^-\to\ell^-\ks$ and $\tau^-\to\ell^-\ks\ks$
(where $\ell = e$ or $\mu$ )
are 
{enhanced}
{in {supersymmetric} 
and many other
{models~\cite{cite:amon,cite:six_fremionic,cite:rpv,cite:rpv2,cite:maria,cite:unparticle}}}.
{Some of these models predict branching fractions
which, for
certain combinations of model parameters,
can be as high as $10^{-7}$;
this 
{level}
is
already accessible
in
high-statistics
{$B$-factory} experiments.
Previously, we obtained
90\% confidence level (C.L.) upper limits
for {the}
$\tau^-\to\ell^-\ks$ 
branching fractions ($\cal{B}$)
using 281 fb${}^{-1}$ of data;
the results were
{${\cal{B}}(\tau^-\rightarrow e^-\ks) < 5.6\times 10^{-8}$
and
${\cal{B}}(\tau^-\rightarrow \mu^-\ks) < 4.9\times 10^{-8}$}~\cite{cite:belle_lks}.
{The BaBar collaboration
has
recently
obtained 90\% C.L.
upper limits
{of}
{${\cal{B}}(\tau^-\rightarrow e^-\ks) < 3.3\times 10^{-8}$
and
${\cal{B}}(\tau^-\rightarrow \mu^-\ks) < 4.0\times 10^{-8}$}
using a data sample of 
469 fb${}^{-1}$~\cite{cite:babar_lks}.}
The 
{most restrictive existing upper limits}
${\cal{B}}(\tau^-\rightarrow e^-\ks\ks) < 2.2\times 10^{-6}$
and 
${\cal{B}}(\tau^-\rightarrow \mu^-\ks\ks) < 3.4\times 10^{-6}$
at the 90\% {C.L.}
{were set
by  the CLEO experiment}
using 13.9 fb${}^{-1}$ of data~\cite{cite:cleo_lksks}.
In this paper,
we {{present} 
a  search for
{the {lepton-flavor-violating}  decays}
$\tau^-\rightarrow \ell^-\ks$
and $\ell^-\ks\ks$}
($\ell = e \mbox{ or } \mu$)~\footnotemark[2]
{using  671 fb$^{-1}$ of data 
collected at the $\Upsilon(4S)$ resonance
and 60 MeV {below}}
with the Belle detector at the 
{KEKB asymmetric-energy $e^+e^-$ collider}~\cite{kekb}. 
\footnotetext[2]{Unless otherwise stated,
{charge-conjugate} decays are 
{included}
throughout
this paper.}

The Belle detector is a large-solid-angle magnetic spectrometer that
consists of a silicon vertex detector (SVD), 
a 50-layer central drift chamber (CDC), 
an array of aerogel threshold 
{{C}herenkov} counters (ACC), a barrel-like arrangement of 
time-of-flight scintillation counters (TOF), and an electromagnetic calorimeter 
comprised of  
CsI(Tl) {crystals (ECL), all located} inside
a superconducting solenoid coil
that provides a 1.5~T magnetic field.  
An iron flux-return located outside of the coil is instrumented to detect $K_{\rm{L}}^0$ mesons 
and to identify muons (KLM).  
The detector is described in detail elsewhere~\cite{Belle}.

{Leptons} are identified
using likelihood ratios
calculated from
the {responses of
various detector subsystems}.
{For electron identification,
the likelihood ratio is defined as
{${\cal P}(e) = {\cal{L}}_e/({\cal{L}}_e+{\cal{L}}_x)$,}
where  ${\cal{L}}_e$ and ${\cal{L}}_x$ are the likelihoods
for electron and non-electron {hypotheses,} 
respectively,
determined using
the ratio of the energy deposit in the ECL to
the momentum measured in the SVD and CDC,
the shower shape in the ECL,
the matching between the position
of {the} charged track trajectory and the cluster position in
the ECL,
the hit information from the {ACC,}
and
the $dE/dx$ information in the CDC~\cite{EID}.
For muon  identification,
the likelihood ratio is defined as
{${\cal P}(\mu) = {\cal{L}_\mu}/({\cal{L}}_{\mu}+{\cal{L}}_{\pi}+{\cal{L}}_{K})$,}
where  ${\cal{L}}_{\mu}$, ${\cal{L}}_\pi$  and ${\cal{L}}_K$ are the likelihoods
for {the}
muon, pion and kaon {hypotheses,} respectively,
based on the matching quality and penetration depth of
associated hits in the KLM~\cite{MUID}.}
{For this measurement,}
{we use hadron identification likelihood variables}
based on
the hit information from the ACC,
{the $dE/dx$ information in the CDC,}
and {the {particle} time-of-flight} from the TOF.
To distinguish hadron species,
we use likelihood ratios,
${\cal{P}}(i/j) = {\cal{L}}_i/({\cal{L}}_i + {\cal{L}}_{j})$,
where ${\cal{L}}_{i}$ (${\cal{L}}_{j}$)
is the likelihood for the detector response
to {a} track with flavor hypothesis $i$ ($j$).

{In order to optimize the event selection
and estimate
the signal efficiency,}
we use {Monte Carlo} (MC) samples.
The signal and background events from generic $\tau^+\tau^-$ decays are
generated by KKMC/TAUOLA~\cite{KKMC}.
The signal MC samples are generated by KKMC
assuming a phase space model for 
{the} $\tau$ decay.
Other {backgrounds,} including
$B\bar{B}$ and 
{continuum}
$e^+e^-\to q\bar{q}$ ($q=u,d,s,c$) events,
Bhabha events,
and 
{two-photon processes,} are generated by
EvtGen~\cite{evtgen},
BHLUMI~\cite{BHLUMI},
and
{AAFH~\cite{AAFH}}, respectively.
{The Belle detector response is simulated 
by a GEANT 3~\cite{cite:geant3}
based program.}
{The event selection is optimized {mode-by-mode}
since the {backgrounds} are mode dependent.}
All kinematic variables are calculated in the laboratory frame
unless otherwise specified.
In particular,
variables
calculated in the $e^+e^-$ center-of-mass (CM) system
are indicated by the superscript ``CM''.

\section{Data Analysis}

{We search for $\tau^+\tau^-$ events,
in which one $\tau$ (signal side) decays
into $\ell\ks$  
{or} $\ell\ks\ks$,
while the other $\tau$ (tag side) decays 
into 
{a final state with}
one charged track, any number of 
additional photons and neutrinos.
We reconstruct 
{each} $\ks$ {meson} candidate from 
{a} $\pi^+\pi^-$ {pair}.}
By selecting 
{decays} into  one charged track 
{on} the tag side,
we reduce background from 
$B\bar{B}$ and $q\bar{q}$ events.
{All charged tracks} and photons 
are required to be reconstructed 
{within a fiducial volume,} 
defined by $-0.866 < \cos\theta < 0.956$,
where $\theta$ is the polar angle with
respect to the direction opposite to the $e^+$ beam.
{We select charged tracks with
{momenta} transverse to the $e^+$ beam
$p_t > 0.1$ GeV/$c$ 
and 
photons with energies
$E_{\gamma} > 0.1$ GeV.}

{Candidate $\tau$-pair events are required to have} 
four {or} six charged tracks 
{with  zero} net charge
for 
{the}
$\ell\ks$ and $\ell\ks\ks$ modes, respectively.
{Events {are} separated into two 
hemispheres corresponding {to} 
the signal 
(three-prong and five-prong 
for {the}
$\ell\ks$ and $\ell\ks\ks$ modes, respectively)
and tag (one-prong) sides
by the plane perpendicular to the thrust
axis~\cite{thrust}. }

We require  one 
{or} two 
$\ks$ candidates
for {the} $\ell\ks$ and $\ell\ks\ks$ modes, respectively.
{The $\ks$ is 
reconstructed 
from two {oppositely charged} 
tracks on the signal side
that {have an invariant mass}   
{0.482 GeV/$c^2 < M_{\pi^+\pi^-} <0.514$ {GeV/$c^2$},
assuming {the} pion {mass} for both tracks.}}}
The $\pi^+\pi^-$ vertex is required to
be displaced from the interaction point (IP)
in the direction of the pion pair momentum~\cite{cite:ks}.
{In order to avoid fake $\ks$ candidates 
from  {photon} conversions
{(i.e.,} 
$\gamma \rightarrow e^+e^-$),}
the invariant mass reconstructed 
by assigning the electron mass to the tracks,
is required to be greater than 0.2 GeV/$c^2$.
{The electron and muon {identification} criteria are}
${\cal P}(e) > 0.9$ with 
{momentum} $p > 0.3$ GeV/$c$
and 
${\cal P}(\mu) > 0.9$ with $p > 0.6$ GeV/$c$,
respectively.
In order to take into account the emission
of  bremsstrahlung photons from the electron,
the momentum of {each}
electron track
is reconstructed by
{adding
the momentum of every photon
within}
0.05 rad 
{of}
{the track {direction}.}
{The electron (muon) identification
efficiency 
for {the} 
$\ell\ks$  modes is 
92\% (87\%) 
and {that} for the $\ell\ks\ks$ modes is
79\% (81\%).}
The difference of efficiencies between 
$\ell\ks$ and $\ell\ks\ks$ 
is due to {the different {signal} 
momentum} {distributions.}
{The probability to misidentify {a} pion
as {an} {electron} and {a} muon}
is below 0.5\% and 3\%, {respectively.}

{In order to suppress background 
from $q\bar{q}$ events,}
the following requirements 
on
the number of the photon candidates on the signal and tag side
($n_{\gamma}^{\rm{SIG}}$ and $n_{\gamma}^{\rm{TAG}}$) {are imposed:}
{$n_{\gamma}^{\rm{SIG}}\leq 1$ and $n_{\gamma}^{\rm{TAG}}\leq 3$.}
For {the} $\ell\ks$ modes only,
we {also}
{require}  {$n_{\gamma}^{\rm{TAG}}\leq 1$}  
{if the  track of the tag side}
is {a} 
lepton 
{to reduce the background, 
{in particular from}
$D^{+}\to\ell^{+}\nu\ks(\to\pi^0\pi^0)$.}

\begin{figure}[t]
\begin{center}
 \resizebox{.35\textwidth}{!}{\includegraphics
 {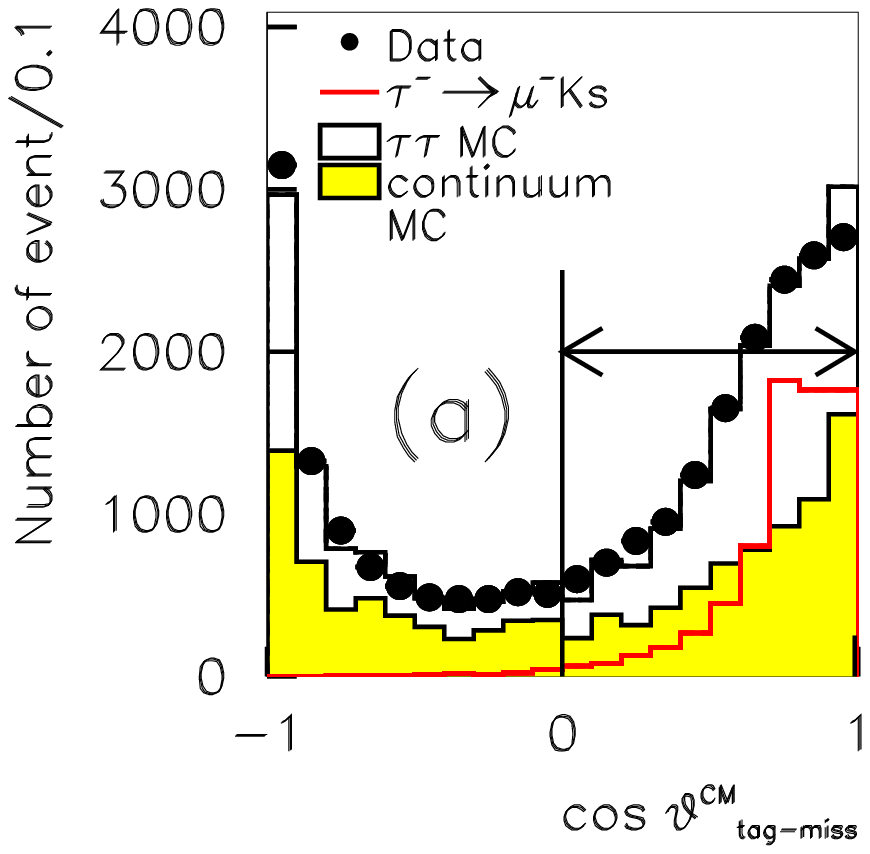}}
\hspace*{-0.7cm}
 \resizebox{.35\textwidth}{!}{\includegraphics
 {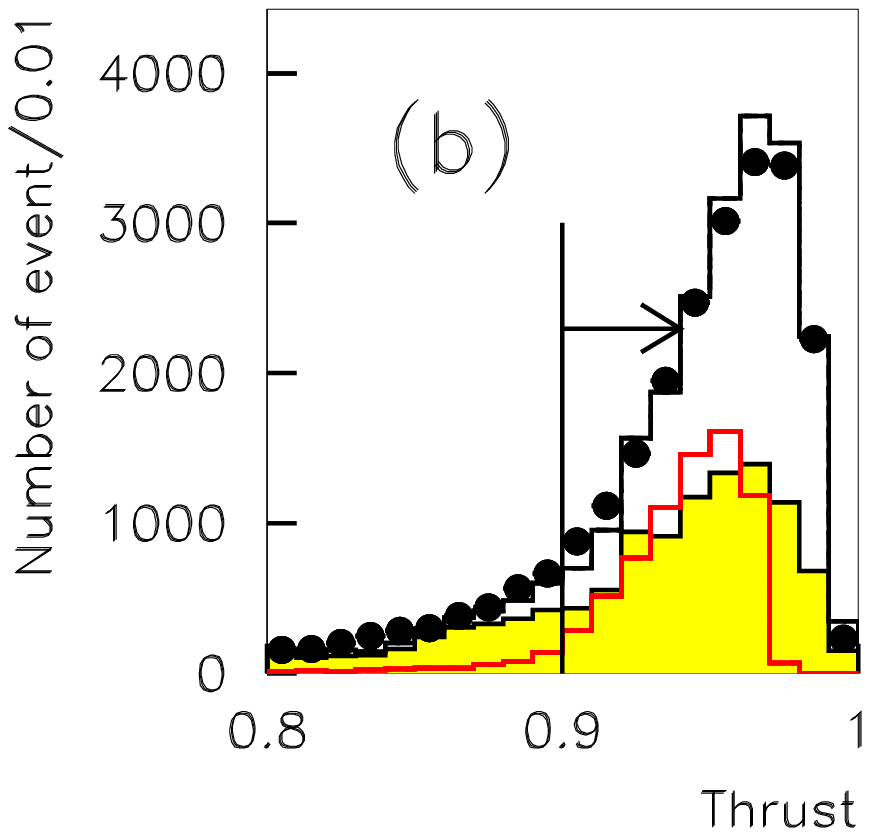}}
\vspace*{-0.1cm}
 \caption{
 {Kinematic distributions used in the event selection
  of {the} $\tau^-\to\mu^-\ks$ mode:
 (a) the cosine of the opening angle between a charged track on the
 tag side and
 {the} 
missing momentum in the CM system ($\cos \theta_{\rm tag-miss}^{\rm CM}$);
 and
 (b) the magnitude of the thrust.
{The signal MC ($\tau^-\to\mu^-\ks$)
 distributions with arbitrary normalization  are shown for
 comparison;
 the background MC
 distributions are normalized to the data luminosity.}
 {Selected regions are indicated
 by {the}
 arrows from the marked cut {boundaries.}}}
}
\label{fig:cut_muks}
 \resizebox{.35\textwidth}{!}{\includegraphics
 {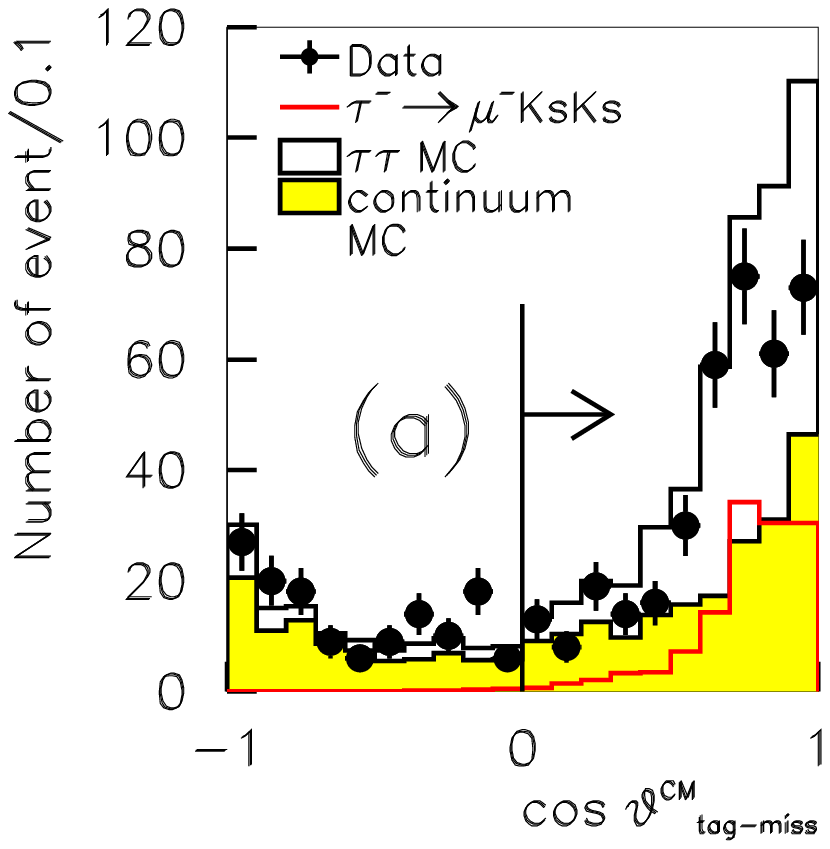}}
\hspace*{-0.7cm}
 \resizebox{.35\textwidth}{!}{\includegraphics
 {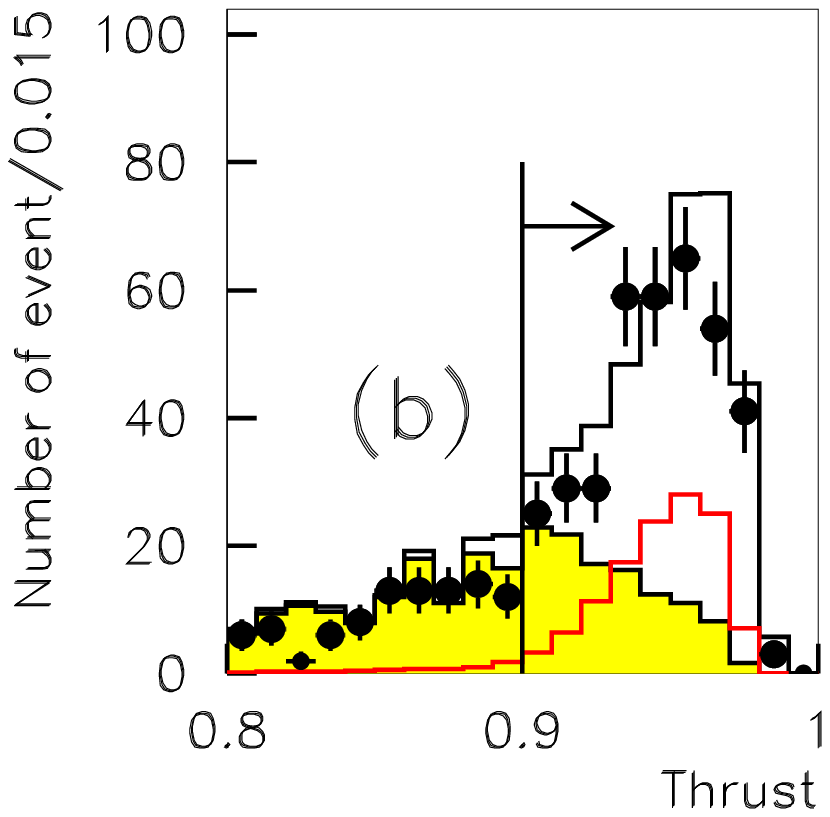}}
\vspace*{-0.1cm}
 \caption{
 {Kinematic distributions used in the event selection 
 of {the}
$\tau^-\to\mu^-\ks\ks$ mode:
 (a) the cosine of the opening angle between a charged track on the
 tag side and
{the}
 missing momentum in the CM system ($\cos \theta_{\rm tag-miss}^{\rm CM}$);
and 
 (b) the magnitude of the thrust.
{The signal MC ($\tau^-\to\mu^-\ks\ks$)
 distributions with arbitrary normalization  are shown for
 comparison;
 the background MC
 distributions are normalized to the data luminosity.}
 {Selected regions are indicated
 by {the}
 arrows from the marked cut {boundaries.}}}
}
\label{fig:cut_muksks}
\end{center}
\end{figure}

To ensure that the missing particles are neutrinos rather
than photons or charged particles 
{that fall outside the detector acceptance,} 
we impose additional requirements on the missing
momentum vector, $\vec{p}_{\rm miss}$, 
calculated by subtracting the
vector sum of the momenta
of all tracks and photons 
from the sum of the $e^+$ and $e^-$ beam momenta.
We require that the magnitude of $\vec{p}_{\rm miss}$ 
{be} greater than
0.4 GeV/$c$ and that 
{its direction {point} into} 
the fiducial volume of the
detector.
{Since neutrinos are emitted only on the tag side,
the direction of 
{$\vec{p}_{\rm miss}$
should lie within the tag side of the event.}}
The cosine of the
opening angle between 
{$\vec{p}_{\rm miss}$}
and
{the} tag-side track 
in the CM system,
{$\cos \theta^{\mbox{\rm \tiny CM}}_{\rm tag-miss}$,}
should be
{$0.0<\cos \theta^{\mbox{\rm \tiny CM}}_{\rm tag-miss}$
for both modes}
(see {Figs.}~\ref{fig:cut_muks} (a) and \ref{fig:cut_muksks} (a)).
{{For the $\ell\ks$ modes, we also require} that
$\cos \theta^{\mbox{\rm \tiny CM}}_{\rm tag-miss}<0.99$
to reduce} background
from
Bhabha, $\mu^+\mu^-$ and {two-photon events,}
as
radiated photons
from the tag-side track
result in missing momentum
if they overlap with
the ECL
clusters {associated with}
the tag-side track.}

To reject the {$q\bar{q}$} background,
the magnitude of the thrust is required to be larger than 0.9 
(see Fig.~\ref{fig:cut_muks} (b) and \ref{fig:cut_muksks} (b)). 
The
{invariant mass reconstructed
{from the charged track and 
{any photon}}}
{on} the tag side
is required to be less than 
1.0 and 1.777 GeV/$c^2$ 
for {the} $\ell\ks$ and $\ell\ks\ks$ 
{modes,} respectively.
{For {the} $\ell \ks$ modes,
{we impose a kaon 
veto  ${\cal{L}}(K/\pi)<0.6$
if the	track on the tag side is a hadron, to suppress 
$e^+e^-\to q\bar{q}$ background{;} due to 
the conservation of strangeness by the strong
interaction, the $\ks$ in such events is 
{often} accompanied by another
kaon.}}

All kinematic distributions for 
{the} $\ell\ks$ modes
shown in Fig.~\ref{fig:cut_muks}
are
{in}
reasonable agreement {between data and background MC} 
{while those for} the $\ell\ks\ks$ modes
shown in Fig.~\ref{fig:cut_muksks}
{clearly differ.}
This difference between the data and background MC
in Fig.~\ref{fig:cut_muksks}
{originates}
from 
{our poor knowledge of
the branching fractions
${\cal{B}}(\tau^-\to\pi^-\ks\ks\nu_{\tau}) = (2.4 \pm 0.5)\times10^{-4}$
and
${\cal{B}}(\tau^-\to\pi^-K^0\bar{K}^0\pi^0\nu_{\tau}) \times 
({\cal{B}}(K^0\to\ks))^2
= 
(3.1 \pm2.3)\times10^{-4}\times1/4$
and {the} 
dynamics of these decays~\cite{PDG}.}
Since the final estimate of the background uses
information {from the data,}
this discrepancy does not directly affect our results.}

\begin{figure}[t]
\begin{center}
 \resizebox{.7\textwidth}{!}{\includegraphics
 {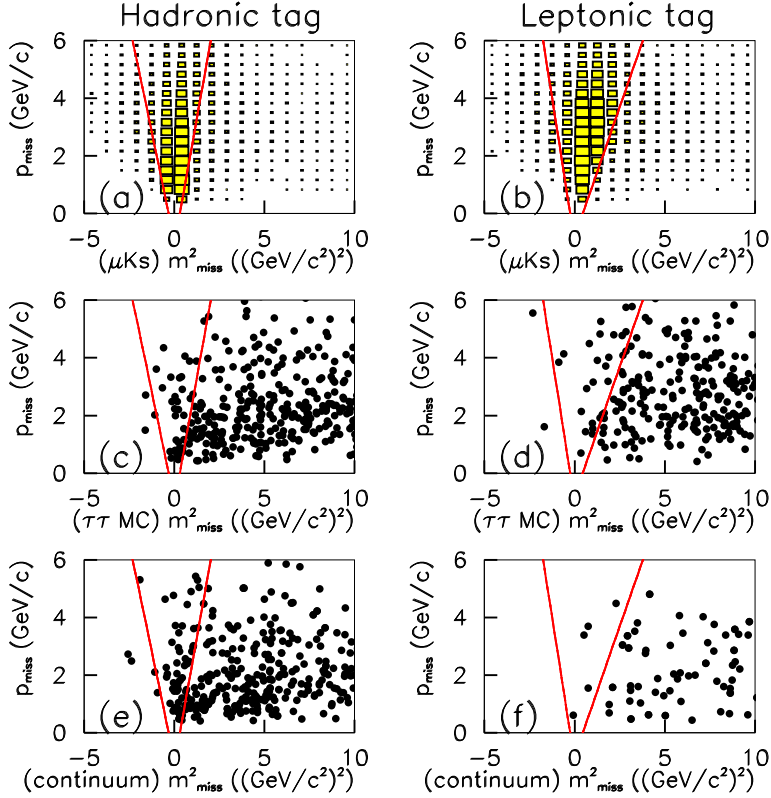}}
 \caption{
Scatter-plots of
{$p_{\rm miss}$
{vs.}
$m_{\rm miss}^2$:
(a), (c) and (e)
show
the signal MC
($\tau^-\to\mu^-\ks$),
the generic $\tau^+\tau^-$ MC
and $q\bar{q}$
distributions,
respectively,
for the hadronic tags
while (b), (d) and (f) show
the same distributions
for the {leptonic tags.}}
Selected regions are indicated by lines.}
\label{fig:pmiss_vs_mmiss2_muks}
\end{center}
\end{figure}

Finally,
to suppress  backgrounds from generic
$\tau^+\tau^-$ and $q\bar{q}$ events,
we apply a selection based on the magnitude of the missing momentum
${p}_{\rm{miss}}$ and {the}
missing mass squared $m^2_{\rm{miss}}$. 
{The latter is defined as
$E^2_{\rm miss}-p^2_{\rm miss}$,
where $E_{\rm miss} = E_{\rm total}-E_{\rm vis}$,
$E_{\rm total}$ is the sum of the beam energies
and $E_{\rm vis}$ is the total visible energy.}
{{We apply different selection criteria depending on the 
{type of one-prong tag:}
the number of
emitted neutrinos is two if the {tagging} 
track is an electron or muon
(leptonic tag) while it is one if the {tagging} 
track is a hadron (hadronic tag).}
{The requirements are}
{{listed} in Table~\ref{tbl:miss_m}}
{(see also Fig.~\ref{fig:pmiss_vs_mmiss2_muks}). }}

\begin{table}
\caption{
The selection criteria {for}
the missing momentum ($p_{\rm{miss}}$) and
missing mass squared ($m^2_{\rm miss}$)
{correlations,}
where
{$p_{{\rm miss}}$} is in GeV/$c$
and  $m^2_{\rm miss}$ is in $(\rm{GeV}/c^2)^2$.
}
\begin{tabular}{c|c|c}\hline\hline
Modes & Hadronic tag & Leptonic tag \\ \hline
$\ell\ks$ & $p_{\rm{miss}} > -3.0\times m^2_{\rm{miss}}-0.9$ 
         & $p_{\rm{miss}} > -4\times m^2_{\rm{miss}}-1.0$ \\
         & $p_{\rm{miss}} > 3.5\times m^2_{\rm{miss}}-1.1$ 
         & $p_{\rm{miss}} > 1.8\times m^2_{\rm{miss}}-0.8$ \\\hline

$\ell\ks\ks$ & $p_{\rm{miss}} > -2\times m^2_{\rm{miss}}-1.0$ 
             & $p_{\rm{miss}} > -2\times m^2_{\rm{miss}}-1.0$ \\
             & $p_{\rm{miss}} > 2\times m^2_{\rm{miss}}-1.0$ 
            & $p_{\rm{miss}} > 1.3\times m^2_{\rm{miss}}-0.8$ \\
\hline\hline
\end{tabular}
\label{tbl:miss_m}
\end{table}

\section{Results}

Signal candidates are examined in 
{two-dimensional plots} 
of the {$\ell\ks$ and $\ell\ks\ks$}  invariant 
mass, $M_{{\rm sig}}$ ($= M_{\ell\ks}$, $M_{\ell\ks\ks}$), 
and the difference of their energy from the 
beam energy in the CM system, {$\Delta E$.} 
A signal event should have $M_{{\rm sig}}$
close to the $\tau$-lepton mass
and 
$\Delta E$ close to 
{zero.}
{For both modes,
the $M_{{\rm sig}}$ and $\Delta E$  resolutions} are parameterized 
from the MC distributions around the peak region 
{using  asymmetric Gaussian  shapes} 
to  {take into account} initial state radiation.
The 
{widths} of these Gaussians 
are shown in Table~\ref{tbl:reso}.

\begin{figure}
\begin{center}
 \resizebox{0.35\textwidth}{0.35\textwidth}{\includegraphics
 {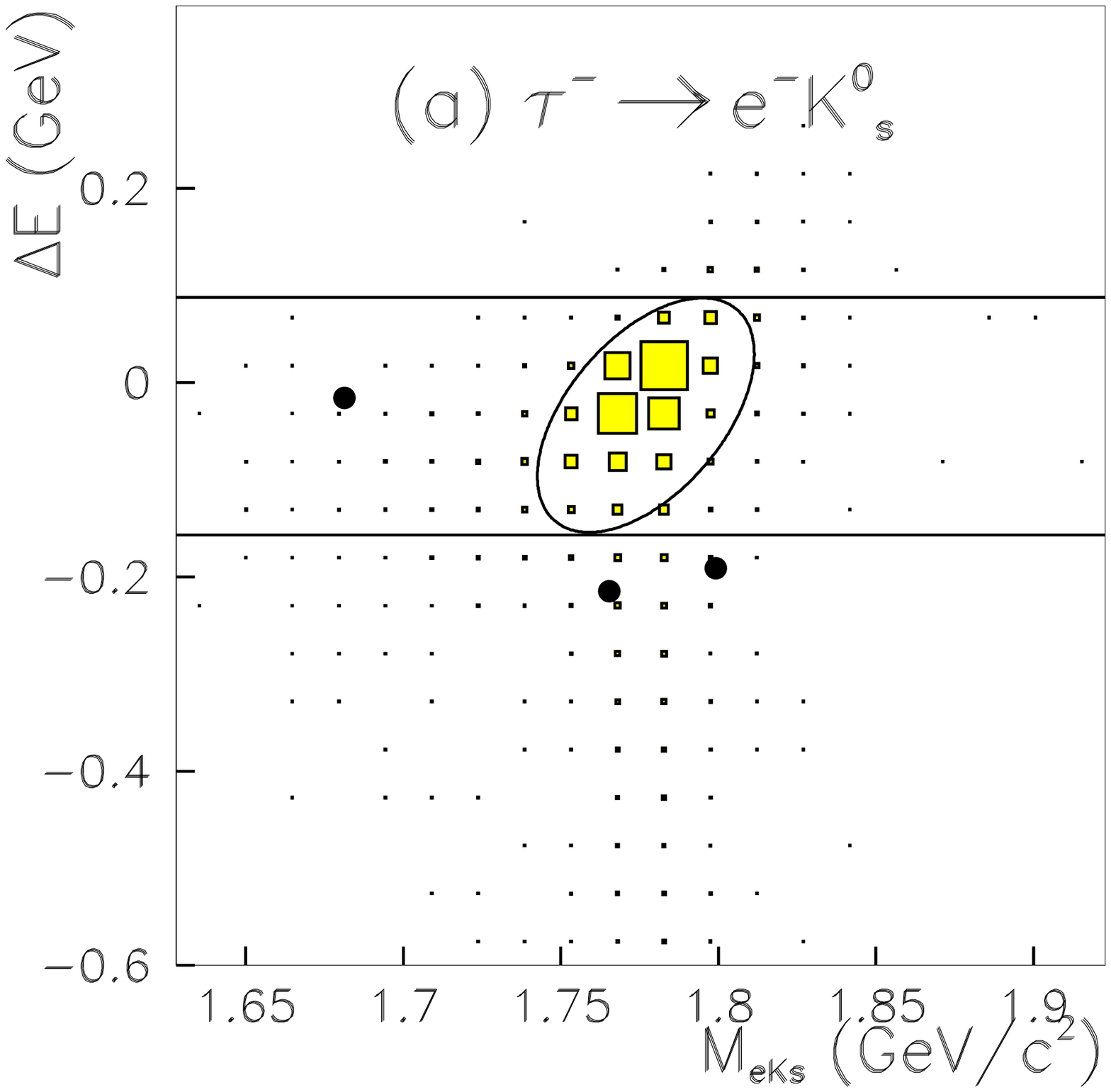}}
 \resizebox{0.35\textwidth}{0.35\textwidth}{\includegraphics
 {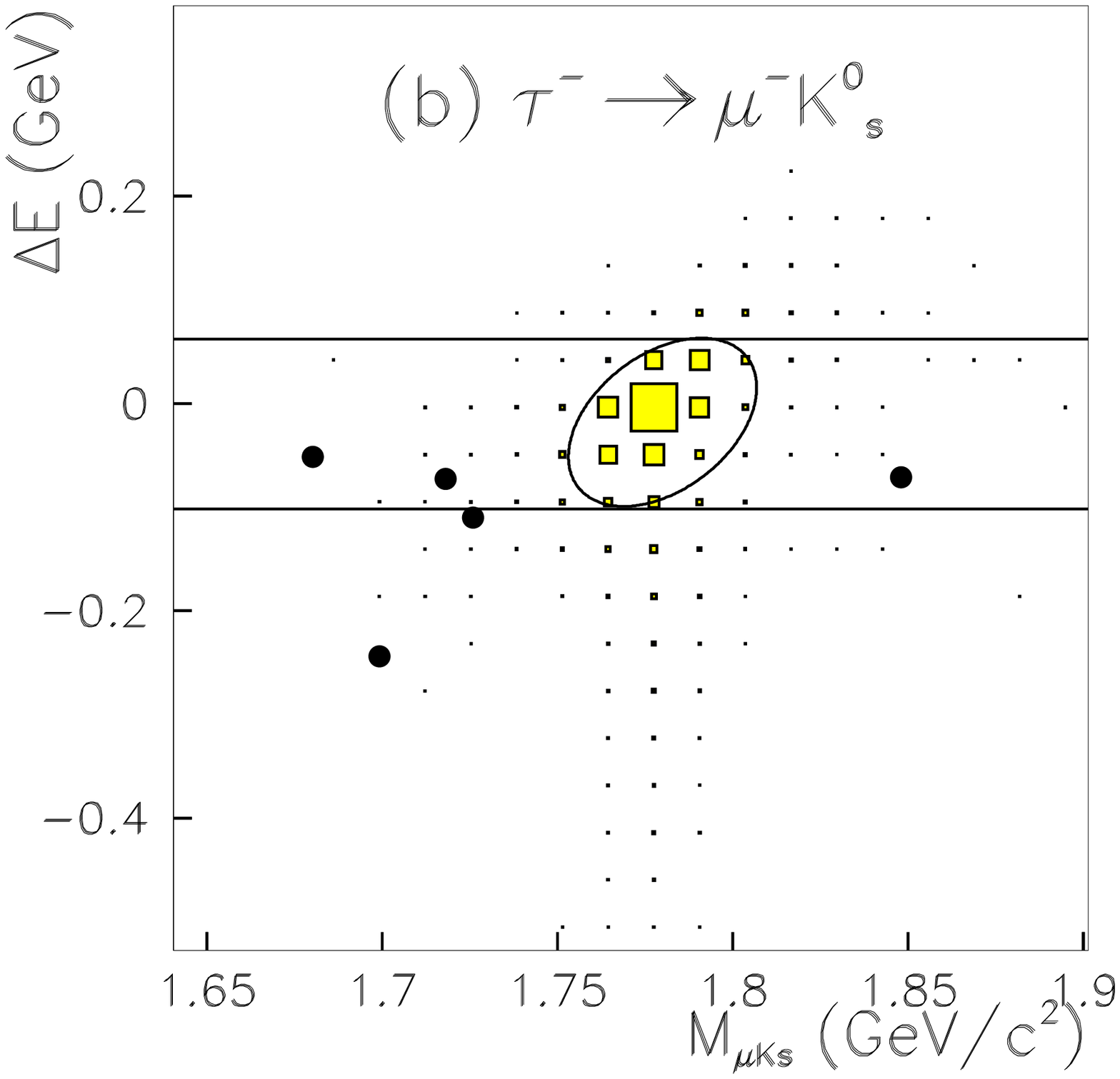}}
 \resizebox{0.35\textwidth}{0.35\textwidth}{\includegraphics
 {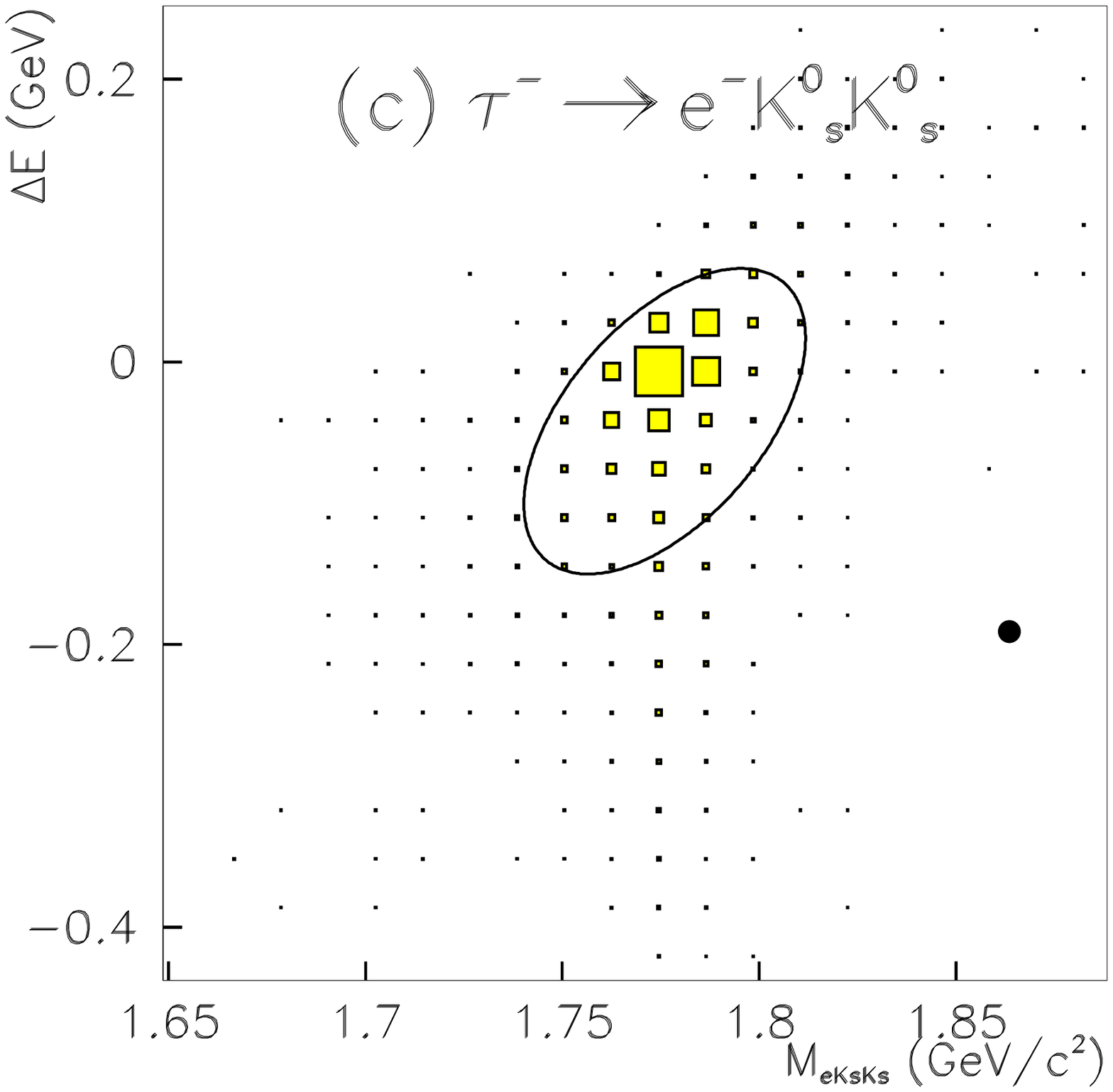}}
 \resizebox{0.35\textwidth}{0.35\textwidth}{\includegraphics
 {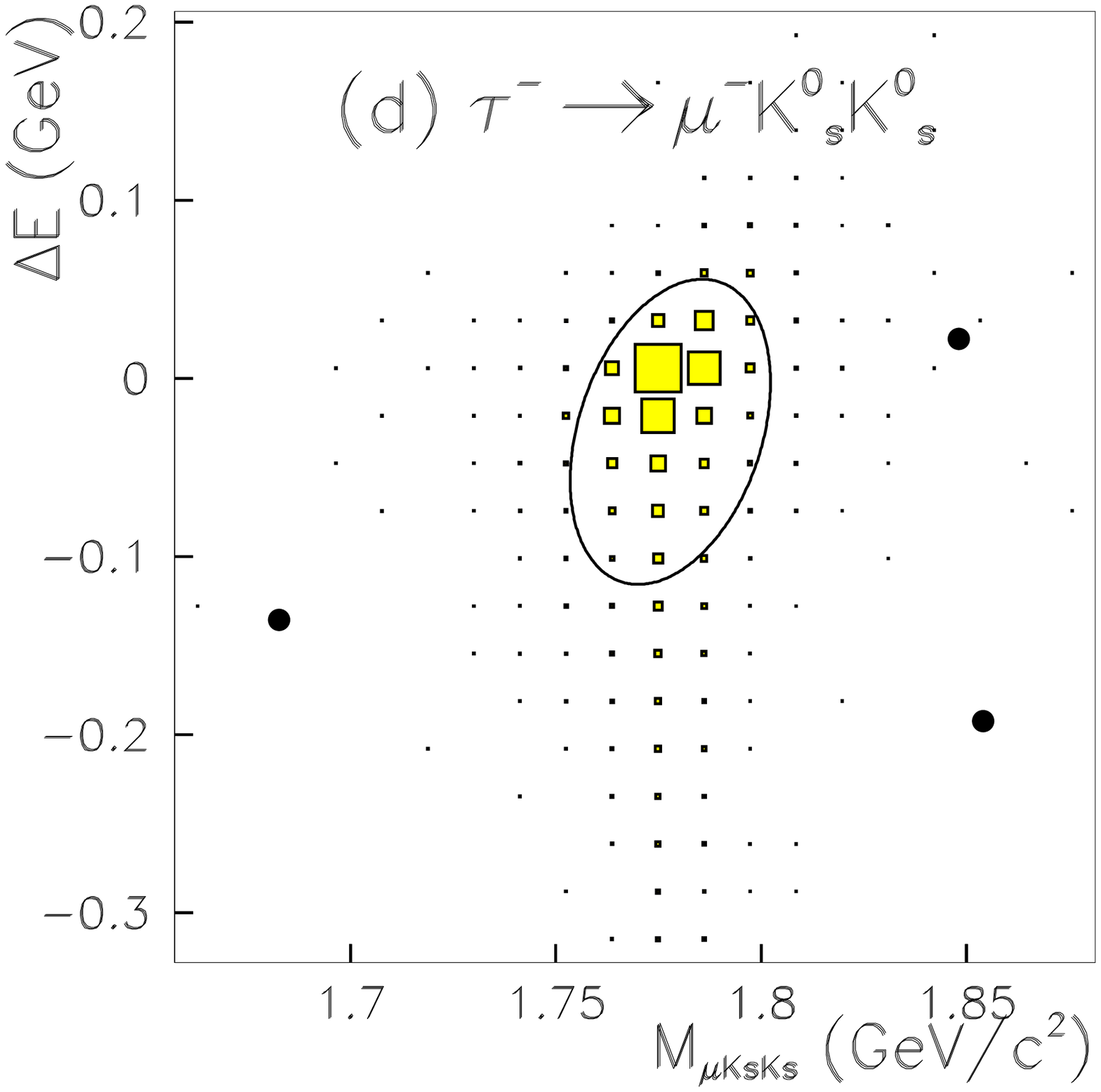}}
\caption{{{
Scatter-plots of data in the 
$M_{\rm sig}$ -- $\Delta{E}$ plane: 
(a), (b), (c) and (d) correspond to
the $\pm 20 \sigma$ area for
the 
$\tau^-\rightarrow e^-\ks$, 
$\tau^-\rightarrow \mu^-\ks$, 
$\tau^-\rightarrow e^-\ks\ks$ and 
$\tau^-\rightarrow \mu^-\ks\ks$ modes, respectively.
{Data is} indicated by the solid circles.
The filled boxes show the MC signal distribution
with arbitrary normalization.
The elliptical signal {regions}} shown by 
{the} solid {curves are} 
used for evaluating the signal yield.
{In (a) and (b),
the region between the horizontal solid lines excluding
the signal region is
used as {a} sideband.}
}}
\label{fig:3}
\end{center}
\end{figure}

To evaluate the branching fractions,
we use 
{an elliptical signal region
that contains} 90\%
of the signal MC events satisfying
{all selection criteria.
The shape of
the signal region is
chosen to minimize its area
and therefore obtain the highest sensitivity.}
We blind the data in the signal region
{until all selection criteria are finalized}
so as not to bias our choice of selection criteria. 
Figure~\ref{fig:3} shows scatter-plots 
for data events and signal MC samples 
distributed over $\pm 20\sigma$ 
in the $M_{\rm sig}-\Delta E$ plane.
{As MC simulation shows, 
the dominant background  in the signal region 
comes from events
with a fake lepton from a pion. 
Therefore, we estimate the 
number of expected background
by multiplying the number of data events in the signal region
with selected hadrons (${\cal{P}}(\ell) \leq 0.9$) by the fake lepton ratio. 
The latter is
calculated as the number of events in the data with 
$P(\ell) > 0.9$ 
divided by
the number of events in the data 
{with $P(\ell) \leq 0.9$} in the sideband region. 
For the $\ell\ks$ modes 
we define {the} sideband 
region as the box inside 
{the two horizontal lines 
(see Fig.~\ref{fig:3} (a) and (b))}
with 
the signal region excluded 
{since 
real leptons from $D^{+}\to\ell^{+}{\nu}\ks$ 
populate 
the region
below  the
$\Delta{E}$ signal one.}
{For the $\ell\ks\ks$ modes,
events that lie within 
{a} $\pm 20\sigma$ region 
but outside the signal region are {treated} 
as sideband events}
(see Fig.~\ref{fig:3} (c) and (d)). 
The final signal efficiency and
the number of expected background
events in {the} signal region for each mode
are summarized in Table~\ref{tbl:eff}.}

\begin{table}
\begin{center}
\caption{
Summary {of} $M_{\rm sig}$ 
and $\Delta E$
{resolutions} ($\sigma^{\rm{high/low}}_{M_{\rm sig}}$ (MeV/$c^2$)
and
$\sigma^{\rm{high/low}}_{\Delta E}$ (MeV)).
Here $\sigma^{\rm high}$ ($\sigma^{\rm low}$)
means the standard deviation
{on the} higher (lower) side of the peak.}
\label{tbl:reso}
\vspace*{0.2cm}
\begin{tabular}{c|cccc} \hline\hline
Mode
& {$\sigma^{\rm{high}}_{M_{\rm{sig}}}$ }
& {$\sigma^{\rm{low}}_{M_{\rm{sig}}}$}
& $\sigma^{\rm{high}}_{\Delta E}$ 
&  $\sigma^{\rm{low}}_{\Delta E}$ 
\\ \hline
$\tau^-\to e^-\ks$ & 7.3  & 7.5 & 19.4 & 30.0 \\
$\tau^-\to \mu^-\ks$ &  6.2& 6.8  & 19.1 & 26.4 \\
$\tau^-\to e^-\ks\ks$ &  5.6 & 6.4  & 12.6 & 21.9 \\
$\tau^-\to \mu^-\ks\ks$ & 5.2  &  6.0 & 11.2 &17.2 \\
\hline\hline
\end{tabular}
\end{center}
\end{table}

\begin{table}
\begin{center}
\caption{ The signal efficiency~($\varepsilon$),
the number of the expected background {events}~($N_{\rm BG}$)
estimated from the  sideband data,
{the} {total}
systematic uncertainty~($\sigma_{\rm syst}$),
{the} number of {observed events}
in the signal region~($N_{\rm obs}$),
90\% C.L. upper limit on the number of signal events including
systematic uncertainties~($s_{90}$)
and 90\% C.L. upper limit on the branching
fraction
for each individual mode. }
\label{tbl:eff}
\begin{tabular}{c|cccccc}\hline \hline
Mode &  $\varepsilon$~{(\%)} &
$N_{\rm BG}$  & $\sigma_{\rm syst}$~{(\%)}
& $N_{\rm obs}$ & $s_{90}$ &
${\cal{B}}~(\times10^{-8})$ \\ \hline
$\tau^-\to e^-\ks$ &  10.2 &
 0.18$\pm$0.18 & 6.6 &
 0 & 2.25 & {$<$}2.6 \\
$\tau^-\to\mu^-\ks$ & 10.7 & 
0.35$\pm$0.21 & 6.8 &
0 & 2.10 &  {$<$}2.3  \\
$\tau^-\to e^-\ks\ks$ &  5.82 &
 0.07$\pm$0.07 & 11.2 &
 0 & 2.44 & {$<$}7.1 \\
$\tau^-\to\mu^-\ks\ks$ & 5.08 & 0.12$\pm{0.08}$ &
 11.3 &
0 &  2.40 & {$<$}8.0 \\
\hline\hline
\end{tabular}
\end{center}
\end{table}

%
% Systematic uncertainties
%

The dominant systematic uncertainties 
on 
{the detection sensitivity} 
{come
from $\ks$ 
{reconstruction}
and
tracking efficiencies.}
{These are 4.5\% per $\ks$ candidate
and 1.0\% per track. 
Other sources of systematic uncertainties
are:
lepton identification (2.2$-$2.7)\%,
MC statistics (0.8$-$1.0)\%,
{trigger efficiency (0.01$-$0.4)\%,
and integrated luminosity (1.4\%).}
{The uncertainty 
{from} 
${\cal{B}}(\ks\to\pi^+\pi^-)$ is negligible.}
All these uncertainties are added in {quadrature}
{to provide total systematic uncertainties 
that range from  6.6\% to 11.3\%.}

{Finally, we 
{examine} the blinded region and find
no data events in the signal region 
{for any of the decay modes}
(see Fig.~\ref{fig:3}). 
{Therefore,} 
we set the following upper limits on the branching fractions
based on the Feldman-Cousins method~\cite{cite:FC}.
{The 90\% C.L. upper limit on the number of signal events
~($s_{90}$) 
{is obtained using the} 
POLE program~\cite{pole},
{based on}
the number of expected {background events}, observed data
and {the systematic uncertainty.}} 
The upper limit on the branching fraction  is then given by
\begin{equation}
{\cal B}(\tau^- \rightarrow \ell^- \ks(\ks)) 
<  \frac{s_{90}}{2 \varepsilon 
{\cal B}( \ks \rightarrow \pi^+\pi^-)^n
N_{\tau\tau}},
\end{equation}
where 
{$\varepsilon$ is the signal efficiency},
{${\cal B}(\ks \rightarrow \pi^+\pi^-) = 
(69.20 \pm 0.5)\%$~\cite{PDG}},
and
$n$ is 1 and 2 for {the}
$\ell\ks$ and $\ell\ks\ks$ modes, respectively.
{The value {$N_{\tau\tau} =  6.17\times 10^8$}} is obtained
from 
{the product of 
{the} integrated luminosity and}
the cross section of {$\tau$-pair production}
{$0.919 \pm 0.003$ nb~\cite{tautaucs}.}}}
The resulting upper limits on the branching fractions 
at the 90\% C.L.
are
\begin{eqnarray*}
&&{\cal B}(\tau^-\rightarrow e^-\ks) < 2.6 \times 10^{-8}, \\
&&{\cal B}(\tau^-\rightarrow \mu^-\ks) < 2.3 \times 10^{-8},\\
&&{\cal B}(\tau^-\rightarrow e^-\ks\ks) < 7.1 \times 10^{-8}, \\
&&{\cal B}(\tau^-\rightarrow \mu^-\ks\ks) < 8.0 \times 10^{-8}.
\end{eqnarray*}
{For the $\ell\ks$ modes,}
{these results 
improve 
{the {existing}
upper limits}
{by about a factor of 2,} 
compared 
{to } our previously published limits~\cite{cite:belle_lks}.
For {the}
$\ell\ks\ks$ modes,
{these results 
improve 
{the upper limits}
by factors of 31 and 43
for {the}
$e\ks\ks$ and $\mu\ks\ks$, respectively,
compared 
{to the} previously published limits obtained 
{by} the {{CLEO}} experiment~\cite{cite:cleo_lksks}.}

\section{Summary}

We have searched for the
{lepton-flavor-{violating}} decays
$\tau^-\rightarrow\ell^-\ks$ and $\ell^-\ks\ks$ ($\ell = e \mbox{ or } \mu$)  
using data collected 
{with} the Belle detector at the KEKB $e^+e^-$ asymmetric-energy collider.
We 
{find} no signal {for {any} decay  modes.}
The following  upper limits on
{branching fractions}
at the 90\% confidence level
are obtained:
${\cal{B}}(\tau^-\rightarrow e^-\ks) < 2.6\times 10^{-8}$, 
${\cal{B}}(\tau^-\rightarrow \mu^- \ks) < 2.3\times 10^{-8}$,
${\cal{B}}(\tau^-\rightarrow e^-\ks\ks) < 7.1\times 10^{-8}$ and
${\cal{B}}(\tau^-\rightarrow \mu^- \ks\ks) < 8.0\times 10^{-8}$. 
{These results are currently the most stringent  
upper limits for the $\ell\ks$ and the $\ell\ks\ks$ modes.}
These limits
{can be used  to}
constrain new physics scenarios beyond the Standard Model.

\section*{Acknowledgments}

% Please paste this acknowledgement into your latex file. 
% updated  1/06/10   Korean section updated, Czech section added
% updated  4/26/09   Korean section updated (add KISTI)
% updated 12/15/08   add Nagoya's TLPRC, 2 Grant-in-Aids (long only)
%                        2 new NNSFC contract no. (long only) 
% updated 11/26/08   Poland: KBN -> MNiSW, Australia: DEST -> DISR
%
%***** Acknowledgments *****
%----------- Long version, for most papers ----------- 
We are grateful to M. Herrero for useful discussions.
We thank the KEKB group for the excellent operation of the
accelerator, the KEK cryogenics group for the efficient
operation of the solenoid, and the KEK computer group and
the National Institute of Informatics for valuable computing
and SINET3 network support.  We acknowledge support from
the Ministry of Education, Culture, Sports, Science, and
Technology (MEXT) of Japan, the Japan Society for the 
Promotion of Science (JSPS), and the Tau-Lepton Physics 
Research Center of Nagoya University; 
the Australian Research Council and the Australian 
Department of Industry, Innovation, Science and Research;
the National Natural Science Foundation of China under
contract No.~10575109, 10775142, 10875115 and 10825524; 
the Ministry of Education, Youth and Sports of the Czech 
Republic under contract No.~LA10033;
the Department of Science and Technology of India; 
the BK21 and WCU program of the Ministry Education Science and
Technology, National Research Foundation of Korea,
and NSDC of the Korea Institute of Science and Technology Information;
the Polish Ministry of Science and Higher Education;
the Ministry of Education and Science of the Russian
Federation and the Russian Federal Agency for Atomic Energy;
the Slovenian Research Agency;  the Swiss
National Science Foundation; the National Science Council
and the Ministry of Education of Taiwan; and the U.S.\
Department of Energy.
This work is supported by a Grant-in-Aid from MEXT for 
Science Research in a Priority Area (``New Development of 
Flavor Physics''), and from JSPS for Creative Scientific 
Research (``Evolution of Tau-lepton Physics'').

\end{document}